\documentclass[aps,pra,a4paper,showpacs,twocolumn]{revtex4}
\usepackage{amssymb}

\usepackage{amsmath}
\usepackage{bm}
\usepackage{graphicx}
\usepackage{ulem}

\begin{document}
\title{Fast SWAP gate by adiabatic passage}
\date{\today}
\author{N. Sangouard}
\email{nicolas.sangouard@u-bourgogne.fr}
\affiliation{Laboratoire de Physique, Universit\'e de Bourgogne, UMR
CNRS 5027, BP 47870,
21078 Dijon Cedex, France}
\author{X. Lacour}
\affiliation{Laboratoire de Physique, Universit\'e de Bourgogne, UMR
CNRS 5027, BP 47870,
21078 Dijon Cedex, France}
\author{S. Gu\'{e}rin}
\affiliation{Laboratoire de Physique, Universit\'e de Bourgogne, UMR
CNRS 5027, BP 47870,
21078 Dijon Cedex, France}
\author{H. R. Jauslin}
\affiliation{Laboratoire de Physique, Universit\'e de Bourgogne, UMR
CNRS 5027, BP 47870,
21078 Dijon Cedex, France}

\begin{abstract}
We present a process for the construction of a SWAP gate which does not require a composition of elementary gates from a universal set. We propose to employ direct techniques adapted to the preparation of this specific gate. The mechanism, based on adiabatic passage, constitutes a decoherence-free method in the sense that spontaneous emission and cavity damping are avoided.
\end{abstract}

\maketitle
\section{Introduction}
The perspective of a high computational power ge\-ne\-rates
intense efforts to build quantum computers. The quantum logical
gates, which are one of the essential building blocks of a quantum
computer, have received a lot of attention. They act on qubits,
whose states ideally should be insensitive to decoherence, easily
prepared and measured. Moreover, the construction of logical gates
requires a robust mechanism with respect to fluctuations of
experimental parameters. The usual approach consists in creating a
set of universal gates \cite{Galindo} such that all logical
quantum gates can, in principle, be obtained from the composition
of gates belonging to this set. The universal sets $\{U_2\}$
\cite{DiVincenzo} and $\{U_1,$ CNOT$\}$ \cite{Barenco} where $U_N$
is a general unitary matrix in $SU(2^N)$ have played a central
role in quantum computation. However, this generic construction
usually requires compositions of many elementary gates. This
entails an accumulation of decoherence and of other detrimental
effects, which become a considerable obstacle for a practical
implementation.

In this paper, we propose a technique to build a fast SWAP gate
obtained by a scheme based on adiabatic passage with an optical
cavity. It does not use the composition of gates but aims instead
at the construction of a specific gate in such a way that losses
and decoherence effects remain as small as possible. This direct
method is faster, i.e. it involves considerable fewer individual
steps than the composition of elementary gates build
independently. It is thus less exposed to losses and decoherence
processes.

We chose a representation of qubits by atomic states driven by
adiabatic fields in a configuration that is particularly
insensitive to decoherence. Indeed, the decoherence due to
spontaneous emission can be avoided if the dynamics follows a dark
state, i.e. a state without components on lossy excited states.
Moreover, the adiabatic principles provide the robustness of the
method with respect to partial knowledge of the model and against
small variations of field parameters. To implement the gates in a
robust manner, one has to control precisely the parameters that
determine the action of the gates. We therefore do not use
dynamical phases, requiring controllable field amplitudes, nor
geometrical phases, requiring a controllable loop in the parameter
space \cite{unanyan,Unanyan1,Duan,Unanyan2}. We use instead static
phase differences of lasers, which can be easily controlled
experimentally.

\begin{figure}[ht!]
{\includegraphics[scale=0.43]{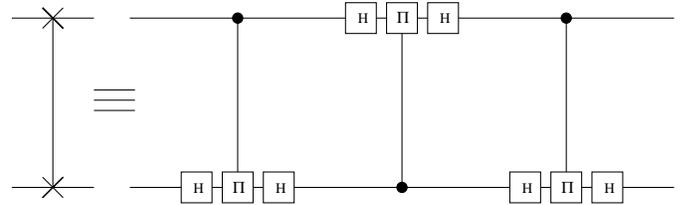}}
\caption{Decomposition of the SWAP gate from C-$phase(\pi)$ gates
and Hadamard (H) gates on the two qubits.}
\label{fig_decomposition}
\end{figure}

In this context of atomic qubits manipulated by adiabatic laser
fields, a mechanism has been proposed in Ref. \cite{kis} to
implement by four pulses all one-qubit gates, i.e. a general
unitary matrix $U_1$ in $SU(2)$ in a tripod system
\cite{unanyan,Unanyan1}. In Ref. \cite{Goto}, five-level atoms are fixed in a single-mode optical cavity and are addressed individually by a set of laser pulses \cite{Pellizari}. The authors proposed sequences of seven pulses to build
a two-qubit controlled-phase gate [C-$phase(\theta)$] and a
two-qubit controlled-NOT gate (CNOT). The configuration of the five-level atoms is defined by adding a second excited state to the
tripod system (see Fig. \ref{scheme_atom}(a)). Since in the tripod, all one-qubit gates can be constructed \cite{kis}, one thus have a mechanism to implement the universal set $\{U_1,$ C-$phase(\theta)\}$ \cite{explanation} from which all
quantum gates can be deduced. For instance, the SWAP gate, which interchanges the
values of two qubits, requires three CNOT gates
\cite{Vatan}, or can be decomposed into six $Hadamard$ gates and
three C-$phase(\pi)$ gates (see Fig. \ref{fig_decomposition}),
which corresponds to at least twenty-one pulses in this system.

Since in the experimental implementation of each gate there are
always losses, due to uncontrolled interactions or decoherence, it
is useful to design direct implementations of specific gates
instead of relegating them to a superposition of many elementary
gates.

\section{System}
We propose an alternative mechanism based on adiabatic passage
along dark states for the construction of the SWAP gate, which
compared to the composition into C-$phase$ and Hadamard gates, or
into CNOT gates of Ref. \cite{Goto}, has the advantage to involve
a much smaller number of pulses and thus to operate in a shorter
time. This mechanism is decoherence-free in the sense that, in the
adiabatic limit and under the condition of a cavity Rabi frequency
much larger than the laser Rabi frequency, the excited atomic
states and the cavity mode are not populated during the dynamics.
We emphasize that the goal here is not to create an alternative
universal set of gates offering the possibility to construct an
arbitrary gate, but to prepare specific logical quantum gates in a
fast way.

\begin{figure}[ht!]
{\includegraphics[scale=0.38]{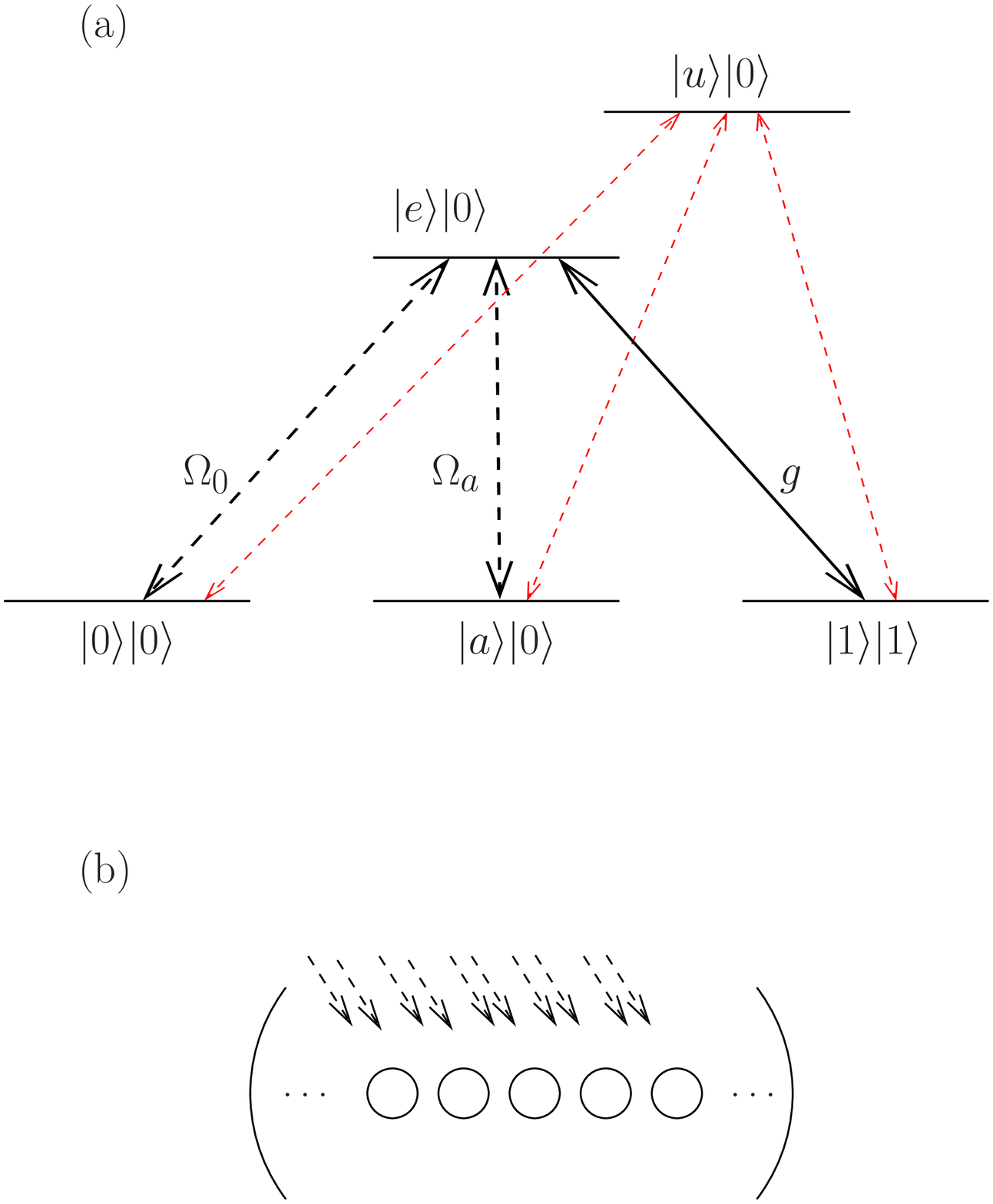}}
\caption{(a) Schematic representation of the five-level atom. Arrows show the laser (full arrows) and cavity (dashed arrows) couplings to perform the swap gate (thick arrows) and the general one qubit gate (thin arrows).
(b) Representation of the atomic register trapped in a single-mode
optical cavity. The atoms are represented by circles, the laser fields by dashed arrows.} \label{scheme_atom}
\end{figure}

We assume that the atoms are fixed inside an optical cavity (Fig.~\ref{scheme_atom}(b)). The proposed mechanism is implemented in the five-level extension of the tripod-type system
in which the universal gate of Refs.~\cite{kis, Goto} can be implemented (Fig.~\ref{scheme_atom}(a)). 

We will use a notation, e.g. in Fig. \ref{scheme_atom} and
\ref{fig_steps}, involving two kets: the left one labels the state
of the atoms (a single or a pair) and the right one the photon
number of the cavity field. The three ground states (for instance
Zeeman levels) $|0\rangle|0\rangle,$ $|a\rangle|0\rangle$ and
$|1\rangle|1\rangle$ are coupled to the excited state
$|e\rangle|0\rangle$ respectively by two lasers associated to the
Rabi frequencies $\Omega_0$ and $\Omega_a$, and by a single mode
cavity associated to the Rabi frequency $g$. The upper state $|u\rangle$ is only used to implement a general one-qubit gate. We assume that the
polarizations and the frequencies are such that each field drives
a unique transition by a one-photon resonant process. As a
consequence, the Stark shifts, which would add detrimental phases,
can be here neglected. (Some estimates are presented below.) The
atomic states $|0\rangle$ and $|1\rangle$ represent the
computational states of the qubit. The ancillary state $|a\rangle$
will be used for the swap operation. The atomic register is fixed
in the single-mode optical cavity (see Fig. \ref{scheme_atom}(a)
(b)). Each atom (labelled by $k$) of the register is driven by a
set of two pulsed laser fields $(\Omega_0^{(k)}(t)$ and
$\Omega_a^{(k)}(t))$ and by the cavity mode $g^{(k)}$ which is
time independent.

\section{Mechanism}
\begin{figure}[ht!]
{\includegraphics[scale=0.38]{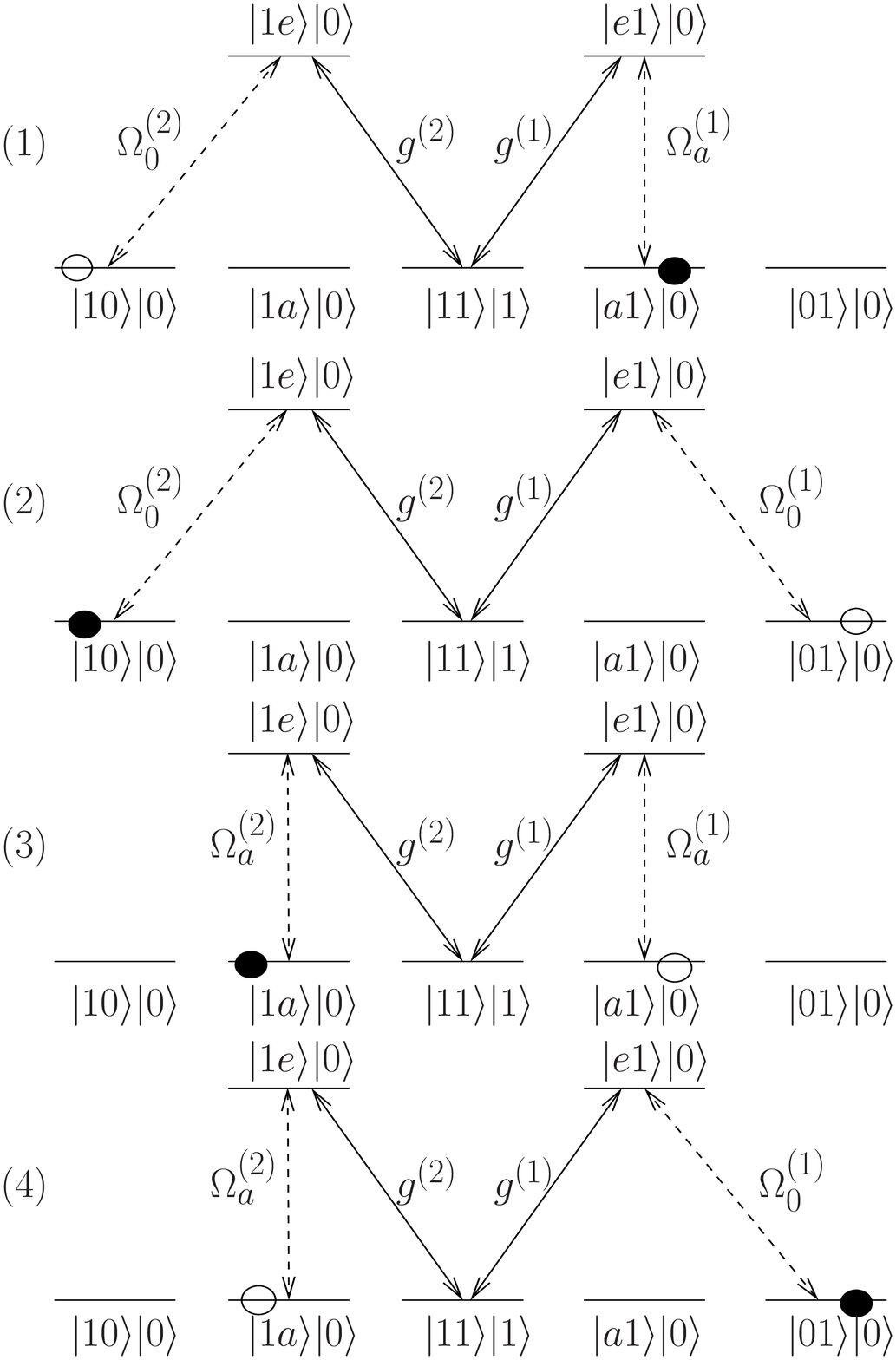}}
\caption{Schematic representation of the four steps of the cons\-truc\-tion of
the SWAP gate. For each step, the intial state is represented by an empty
circle whereas the final state is symbolised by a full black circle.}
\label{fig_steps}
\end{figure}

We propose a mechanism to prepare a SWAP gate
with the help of a simple interaction scheme. The SWAP gate acts
on two qubits as follows. The initial state $|\psi_i\rangle$ of
the atoms in the cavity, before interaction with the lasers, is
defined as
\begin{equation}
\label{init} |\psi_i\rangle= \alpha |00\rangle |0\rangle + \beta
|01\rangle |0\rangle+ \gamma |10\rangle |0\rangle+\delta
|11\rangle |0\rangle,
\end{equation}
where the indices $s_1,s_2$ of the states of the form
$|s_1s_2\rangle|0\rangle$ denote respectively the state of the
first and second atom, and $|0\rangle$ is the initial vacuum state
of the cavity-mode field. $\alpha, \beta, \gamma, \delta$ are
complex coefficients. The swap gate exchanges the values of the two qubits leading to the output state
\begin{equation}
|\psi_o\rangle= \alpha |00\rangle |0\rangle + \gamma |01\rangle
|0\rangle + \beta |10\rangle |0\rangle+\delta |11\rangle
|0\rangle.
\end{equation}
The main idea to construct this gate is represented in Fig. \ref{fig_steps}. It consists of exchanging the values of the qubits by the use of an ancillary ground state.
Adiabatic passage along dark states (i.e. with no components in
the atomic excited states and a negligible component in the
excited cavity states) will be used.

The four steps can be
summarized as follows:

 \underline{Step (1)}: The population of $|10\rangle |0\rangle$ is completely
transferred into
 $|a1\rangle|0\rangle$ by the use of two resonant pulses
 $\Omega_{a}^{(1)}$, $\Omega_{0}^{(2)}$ switched on and off in a
counterintuitive pulse
 sequence (i.e. $\Omega_{a}^{(1)}$ before $\Omega_{0}^{(2)}$.) After the interaction, the state becomes
\begin{equation}
\label{state_step1}
|\psi_1\rangle=\alpha |00\rangle |0\rangle + \beta |01\rangle |0\rangle +
\gamma |a1\rangle |0\rangle+\delta |11\rangle |0\rangle.
\end{equation}

 \underline{Step (2)}: With a similar technique, the population of
$|01\rangle|0\rangle$
 is transferred into $|10\rangle|0\rangle$ by the use of the counterintuitive
sequence of two pulses
 $\Omega_0^{(2)}$, $\Omega_0^{(1)}$ leading to the state
 \begin{equation}
 |\psi_2\rangle=\alpha |00\rangle |0\rangle + \gamma |a1\rangle |0\rangle +
\beta |10\rangle |0\rangle + \delta |11\rangle |0\rangle.
 \end{equation}

 \underline{Step (3)}: The population of $|a1\rangle|0\rangle$ is transferred into
 $|1a\rangle|0\rangle$ by the use of the sequence
 $\Omega_a^{(2)}$, $\Omega_a^{(1)}$ giving
 \begin{equation}
|\psi_3\rangle=\alpha |00\rangle |0\rangle + \gamma |1a\rangle |0\rangle +
\beta |10\rangle |0\rangle + \delta |11\rangle |0\rangle.
 \end{equation}

\begin{figure}[ht!]
{\includegraphics[scale=0.55]{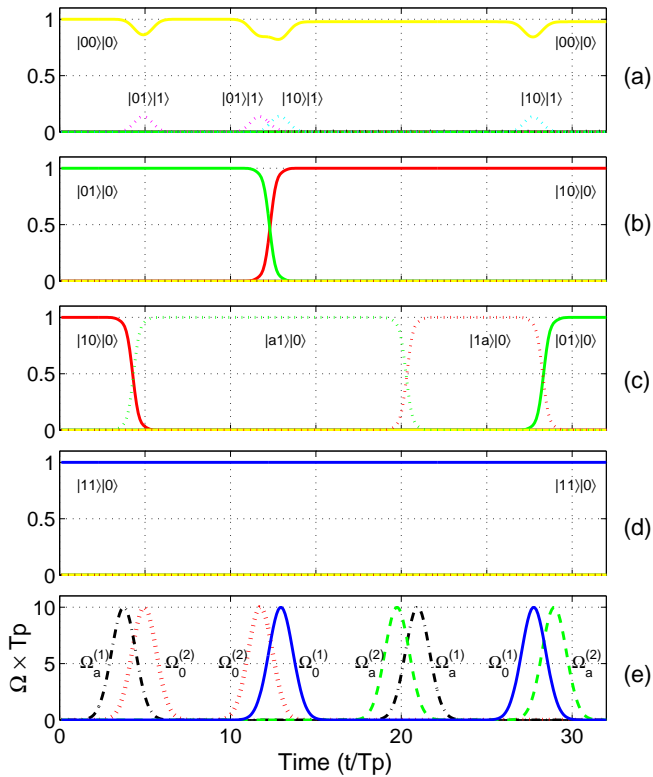}} \caption{(Colour
online) Numerical simulation exhibiting the populations for the
initial conditions: (a) $|00\rangle|0\rangle,$ (b)
$|01\rangle|0\rangle$ (c) $|10\rangle|0\rangle$ and (d)
$|11\rangle|0\rangle.$ The states which are populated during the
interaction with the pulses or between two steps are indicated.
(e) Rabi frequencies. The parameters used are $\Omega_{\max}
T_p=10,$ $g T_p=25.$ The delay between two pulses of the same
steps is $1.2T_p.$} \label{time_evolution}
\end{figure}

 \underline{Step (4)}: The population of $|1a\rangle|0\rangle$ is transferred into
$|01\rangle|0\rangle$
 by the use of the sequence $\Omega_0^{(1)}$, $\Omega_a^{(2)}.$ As a result,
the system is
 in the state
 \begin{equation}
|\psi_4\rangle=\alpha |00\rangle |0\rangle + \gamma |01\rangle
|0\rangle + \beta |10\rangle |0\rangle+\delta |11\rangle
|0\rangle,
 \end{equation}
which coincides with the output state of the SWAP gate. \\
In what follows, we give the instantaneous eigenvectors connected with the initial condition and that are thus adiabatically followed by the dynamics for the four steps. We show that they are associated to dark states with no component in the atomic excited states and a negligible component in the excited cavity states.\\
Since the lasers do not couple the atomic state $|1\rangle$, the
state $|11\rangle|0\rangle$ of the initial
condition (\ref{init}) is decoupled from the other ones. The other
states of (\ref{init}) are connected to two orthogonal decoupled
subspaces denoted ${\cal H}_{7}$ and ${\cal H}_{16}$ of dimension $7$ and $16$ respectively.
For each step, one ground state $|0\rangle$ or $|a\rangle$ of each atom is coupled by
a laser field to the excited state, and the other one is not
coupled to the excited state. To summarize the calculation of the
instantaneous eigenstates for the four steps, we introduce the
following notation : the state coupled by a laser field is labeled
$|{L}^{(i)}\rangle$ ($|0^{(i)}\rangle$ or
$|a^{(i)}\rangle$) and the non-coupled state
$|{N}^{(i)}\rangle$ ($|a^{(i)}\rangle$ or
$|0^{(i)}\rangle$). The index $i=1,2$ labels the atom $i$.
The instantaneous eigenstates in each subspace ${\cal H}_7$ and ${\cal H}_{16}$ can be characterised as follows: in ${\cal H}_7$, the states
$|{N}^{(1)}1\rangle |0\rangle$ and
$|1{N}^{(2)}\rangle |0\rangle$ are not coupled by the lasers and thus do not participate
to the dynamics. Only the atomic dark state (i.e. without
component in the excited atomic states) \cite{Pellizari}:
\begin{eqnarray}
\label{dark3} |\phi_{7}\rangle&\propto&g^{(1)}\Omega^{(2)}
|{L}^{(1)}1\rangle|0\rangle+
g^{(2)}\Omega^{(1)}|1{L}^{(2)}\rangle
|0\rangle
\nonumber\\
&& -\Omega^{(1)}\Omega^{(2)}|11\rangle |1\rangle,
\end{eqnarray}
(where the normalisation coefficient has been omitted)
participates to the dynamics. The first step, associated to
${L}^{(1)}\equiv a$,
${L}^{(2)}\equiv 0$,
$\Omega^{(1)}\equiv\Omega_a^{(1)}$,
$\Omega^{(2)}\equiv\Omega_0^{(2)}$ leads to the initial and final
connections symbolically written as
$|10\rangle|0\rangle\rightarrow|\phi_{7}\rangle\rightarrow|a1\rangle|0\rangle$
(see Fig. 3). The second, third, and fourth steps give
respectively the connections
$|01\rangle|0\rangle\rightarrow|\phi_{7}\rangle\rightarrow|10\rangle|0\rangle$,
$|a1\rangle|0\rangle\rightarrow|\phi_{7}\rangle\rightarrow|1a\rangle|0\rangle$,
and
$|1a\rangle|0\rangle\rightarrow|\phi_{7}\rangle\rightarrow|01\rangle|0\rangle$.
We determine four atomic dark states in the subspace ${\cal
H}_{16}$ connected to the state $|00\rangle|0\rangle$ of the
initial condition (\ref{init}):
\begin{subequations}
\begin{eqnarray}\label{dark_1}
|\phi_{16(1)}\rangle&\propto&\Omega^{(2)}
|{N}^{(1)}1\rangle|1\rangle-g^{(2)}|{N}^{(1)}
{L}^{(2)}\rangle
|0\rangle\\
|\phi_{16(2)}\rangle&\propto& g^{(1)}g^{(2)}\sqrt{2}
|{L}^{(1)}{L}^{(2)}\rangle|0\rangle-
g^{(2)}\Omega^{(1)}\sqrt{2}|1{L}^{(2)}\rangle
|1\rangle \nonumber\\
&&-g^{(1)}\Omega^{(2)}\sqrt{2}|{L}^{(1)}1\rangle
|1\rangle+ \Omega^{(1)}\Omega^{(2)}|11\rangle |2\rangle\\
|\phi_{16(3)}\rangle&=&|{N}^{(1)}{N}^{(2)}\rangle
|0\rangle,\\
 |\phi_{16(4)}\rangle&\propto& \Omega^{(1)}
|1{N}^{(2)}\rangle|1\rangle-
g^{(1)}|{L}^{(1)}{N}^{(2)}\rangle
|0\rangle.
\end{eqnarray}
\end{subequations}
The state $|00\rangle|0\rangle$ is connected initially and finally
to the dark state $|\phi_{16(n)}\rangle$ at the $n^{\text{th}}$
step. The phase term of the final state is equal to one: (i) the optical phase is null since the populated atomic states are degenerate, (ii) the dynamical phase is reduced to zero since the eigenvalues associated to each dark state are null and (iii) the geometric phase is equal to zero. Indeed, at every time $\langle\phi^{\prime}_d(s)|\frac{d}{ds}|\phi_d(s)\rangle=0$ ($|\phi^{\prime}_d(s)\rangle$ and $|\phi_d(s)\rangle$ being two different or identical dark eigenstates) since for $|\phi^{\prime}_d(s)\rangle=|\phi_d(s)\rangle,$ the phase of the lasers is constant during each step and for $|\phi^{\prime}_d(s)\rangle \neq |\phi_d(s)\rangle,$ the dark
states belong to orthogonal subspaces. \\
Since the dynamics follows atomic dark
states, the excited atomic state is never populated (in the
adiabatic limit). Moreover, the projections of the dark states into
the excited cavity photon states can be made negligible if $g^{(i)} \gg
\Omega^{(i)}$ \cite{malinovky}. In this case, the mechanism we
propose is a decoherence-free method in the sense that the process
is not sensitive to spontaneous emission from the atomic excited
states and to the lifetime of photons in the optical cavity.\\
We present the numerical validation of the mechanism proposed for
the construction of the SWAP gate. We show in Fig.
\ref{time_evolution}, the time evolution of four initial states:
in (a) and (d) the population of the initial states
$|00\rangle|0\rangle$ and $|11\rangle|0\rangle$ respectively stays
in these states after the interaction with the eight pulses, in
(b) and (c) the population of the initial states
$|01\rangle|0\rangle$ and $|10\rangle|0\rangle$ are exchanged. In
(e), we show the Rabi frequencies associated to each pulses. The
laser Rabi frequencies are all chosen of the form
$\Omega(t)=\Omega_{\max} e^{-\left(\frac{t}{T_p}\right)^2}.$ Since
the four steps of the mechanism can be seen as double-STIRAPs
\cite{bergmann1,bergmann2}, each STIRAP involving one laser and
the cavity, the amplitudes of the coupling must satisfy
$\Omega_{\max} T_p, gT_p\gg 1$ to fulfill the adiabatic
conditions. The delay between two pulses of the same step is
chosen equal to $2 \times 0.6T_p$ to minimize the non-adiabatic
losses \cite{vitanov}. Moreover, the condition $g \gg
\Omega_{\max}$ guarantees that the cavity mode is negligibly
populated during the interaction with the pulses.

\section{Discussion}
We notice that the two identical pulses $\Omega_0^{(2)}$ used
successively in the first and second step can be replaced by a
single pulse. The process we propose then requires the use of only
seven adiabatic pulses. \\
As a realistic atomic level scheme, we consider the
2$^3S_1$- 2$^3P_0$ transition in metastable Helium, of linewidth
$\Gamma=10^7$ s$^{-1}$ (see for instance \cite{Lawall}). The Rabi
frequencies are $\Omega\sim 10^8\sqrt{I}$ s$^{-1}$ with the field
intensities $I$ in W/cm$^2$. We obtain beyond the resonant
approximation using the polarizability an upper estimate for the
Stark shifts (in absolute value) as $S\sim 100 I$ s$^{-1}$. Taking
into account the loss of the intermediate state requires
$(\Omega_{\max}T_p)^2\gg\Gamma T_p$ for adiabatic passage, which
is satisfied for $\Omega_{\max}T_p\gg1$ and
$\Omega_{\max}\gg\Gamma$. We use $\Omega_{\max}T_p=10$
corresponding to e.g. $I=10^4$ W/cm$^2$ and $T_p=1$ ns, which
gives $\Omega_{\max}\sim10^{10}$ s$^{-1}$ and an estimate of the
phases $S_{\max}T_p\sim10^{-3}\ll2\pi$. They can therefore be
neglected.\\
The swap method we have presented can be extended to build a CNOT
gate. The process is composed of six steps: We first transfer the
population of the state $|1\rangle$ of the second atom in the
ancillary state $|a\rangle$ by STIRAP using two additional
resonant laser field with the upper state $|u\rangle.$ The next
four steps allow to interchange the populations of the states
$|10\rangle|0\rangle$ and $|1a\rangle|0\rangle$ by a similar swap
method that the one shown in this paper. The last step transfers
back the population of the ancillary state $|a\rangle$ of the
second atom in the state $|1\rangle.$ The population
transfers are realised by adiabatic passage along dark states. We
thus obtain a direct and decoherence-free method for the creation
of the CNOT gate that requires the use of eleven pulses. In
comparison to the method proposed in Ref. \cite{Goto}, this
technique does not use f-STIRAP (in which the ratio of two pulses
has to be controlled \cite{vitanov}). A specific system (such as
Zeeman states) is not necessarily required to guarantee the robustness of the technique.

\section{Conclusion}
In conclusion, we have proposed to use a mechanism adapted to the
construction of a specific gate instead of relying on compositions
of a large number of elementary gates. We have illustrated this
idea by the construction of a SWAP gate in a system where all
one-qubit gates and the C-$phase$ gate can be built. This
technique requires the use of a cavity and seven pulses
in a double-STIRAP configuration instead of twenty one pulses when
the SWAP gate is created from the composition of elementary gates.
It is robust against variations of amplitude and duration of the
pulses and of the delay between the pulses. Moreover, it
constitutes a decohence-free method in the sense that the excited
states with short life-times are not populated and the cavity mode
has no photon during the process. We conclude by noticing that
this technique allow one to entangle the qubits on which it acts
by manipulating the phase of the pulses. In this case, we get the
composition of gates that could offer interesting possibilities to
execute rapidly quantum algorithms in which this composition is
required.

Work is in progress to generalize this fast method to other gates.
This requires other pulse sequences and involves different dark
states.



\begin{references}
\bibitem{explanation} The set $\{U_1, CNOT\}$ is a universal set and the $CNOT$ gate can be obtained by the composition of two Hadamard gates and one C-$phase(\pi)$ gate.
\end{references}
\end{document}